\documentclass[fleqn,10pt]{wlscirep}
\usepackage[utf8]{inputenc}
\usepackage[T1]{fontenc}
\usepackage{comment}
\title{Revisiting the Abundance of Topological Materials}
\author[1,2,*]{Hossein Mirhosseini}
\author[3]{Luis Elcoro}
\author[1,2]{Andreas Kn\"upfer}
\author[1,2]{Thomas D. K\"uhne}
\affil[1]{Center for Advanced Systems Understanding (CASUS), D-02826 G\"orlitz, Germany}
\affil[2]{Helmholtz-Zentrum Dresden-Rossendorf (HZDR), D-01328 Dresden, Germany}
\affil[3]{Department of Physics, Faculty of Science and Technology, University of the Basque Country, Bilbao, 48013, Spain}
\affil[*]{h.mirhosseini@hzdr.de}


\begin{abstract}
The classification of topological materials is revisited 
using advanced computational workflows that integrate hybrid density functional theory calculations with exact Hartree-Fock exchange. Unlike previous studies, our workflow optimizes atomic configurations obtained from the Materials Project Database, followed by precise electronic structure calculations. Our results based on hybrid density functional theory calculations reveal that only 15\% of materials are topologically nontrivial, which is in stark contrast to the previously reported 30\% based on semi-local exchange and correlation functionals. This discrepancy underscores the critical dependence of topological classifications on accurate atomic and electronic structures, rendering the abundance of topological materials much lower than generally assumed. 

\end{abstract}
\begin{document}

\flushbottom
\maketitle

\thispagestyle{empty}

\begin{comment}
\noindent Please note: Abbreviations should be introduced at the first mention in the main text – no abbreviations lists. Suggested structure of main text (not enforced) is provided below.
\end{comment}

\section*{Introduction}
Topological materials (TMs) are distinguished by their unique electronic structures, which give rise to unconventional electromagnetic properties and robust surface states protected by time-reversal symmetry~\cite{Kane2005, Bernevig2006, Fu2007}. These features have garnered considerable interest due to their potential applications in advanced technologies such as quantum computing and spintronics~\cite{He2019, Scappucci2021}. The discovery of new TMs is a rapidly advancing field driven by both theoretical predictions and experimental breakthroughs. However, the identification of TMs has traditionally relied on empirical methods, resulting in a relatively limited number of known TMs. In this context, topological quantum chemistry (TQC) and the theory of symmetry indicators have emerged as highly effective tools that can significantly enhance the efficiency and accuracy of the discovery and classification of TMs~\cite{Bradlyn2017, Elcoro2021, Po2017, Kruthoff2017, Song2018, Khalaf2018}.

Predicting the topological nature of materials using first-principles calculations involves quantum mechanical electronic structure calculations, symmetry analysis, and the determination of topological invariants~\cite{Elcoro2017, Cano2018, Vergniory2017, Bradlyn2018}. In pioneering studies, two independent research groups performed high-throughput calculations based on density functional theory (DFT) to assess the topological nature of the non-magnetic materials available in the Inorganic Crystal Structure Database (ICSD). The investigation of Vergniory \emph{et al.}~\cite{Vergniory2019} revealed that 27\% of investigated materials are topological. Zhang \emph{et al.} found that 30\% of materials available in the ICSD are topologically nontrivial~\cite{Zhang2019}. Extending this approach to magnetic systems, Xu \emph{et al.}~\cite{Xu2020} have conducted a high-throughput search within the Magnetic Materials Database~\cite{Gallego2016a, Gallego2016b}, identifying 130 enforced semimetals and topological insulators among magnetic materials.

DFT calculations can be performed at different levels of accuracy, for example using different exchange-correlation (XC) functionals, providing different degrees of accuracy in capturing the electronic properties. While conventional approximations to the XC functional such as the Generalized Gradient Approximation (GGA) are generally accurate for atomic structure optimization, hybrid functionals are often favored for electronic structure calculations. This preference arises from the improved accuracy of hybrid functionals in describing electronic properties by incorporating a fraction of the exact Hartree-Fock exchange mixed with the exchange from a conventional DFT functional. Hybrid functionals have been demonstrated to enhance the description of localized states, improve band dispersion, and yield more accurate bandgap predictions. However, the impact of hybrid functionals varies across different materials, as not all compounds exhibit significant localization effects. Furthermore, bandgap corrections are most pronounced in strongly correlated systems or materials with inherently small bandgaps, where conventional functionals tend to underestimate electronic interactions~\cite{Heyd2003,Heyd2006,Paier2006,Perdew1996}.

To date, high-throughput workflows for classifying TMs have primarily relied on conventional DFT calculations and experimentally determined atomic structures. While these workflows represent a significant advancement, they also prompt two critical questions: first, how does the number of identified TMs vary when different electronic structure calculation methods are employed? Second, how sensitive is the topological classification of materials to their atomic structures? To address these questions, we have developed workflows for evaluating the topological properties of non-magnetic materials, utilizing the Heyd-Scuseria-Ernzerhof (HSE)~\cite{Heyd2003, Heyd2006} hybrid functional to calculate the electronic structures of optimized atomic configurations obtained from the Materials Project Database (MPDB)~\cite{Anubhav2013}. The specifics of our methodology are detailed in the Methods section.

\section*{Results and Discussion}
The workflow utilized in this study, shown in Figure 1, is based on the methodologies described in Refs.~\cite{Vergniory2019, Zhang2019}, with two key modifications. Specifically, we optimized the atomic structures and computed the electronic structures employing both the PBE and the HSE functionals. In contrast to previous studies, where the atomic structures from the ICSD were used for electronic structure calculations, in our workflow, we optimized the atomic structures retrieved from the MPDB. It is noteworthy that experimentally identified structures often do not represent the optimized atomic configurations from the perspective of DFT calculations. As a result, the electronic structures of experimentally identified atomic configurations frequently differ from those obtained from DFT-optimized structures. 
\begin{figure}[ht]
\centering
\includegraphics[width=0.49\linewidth]{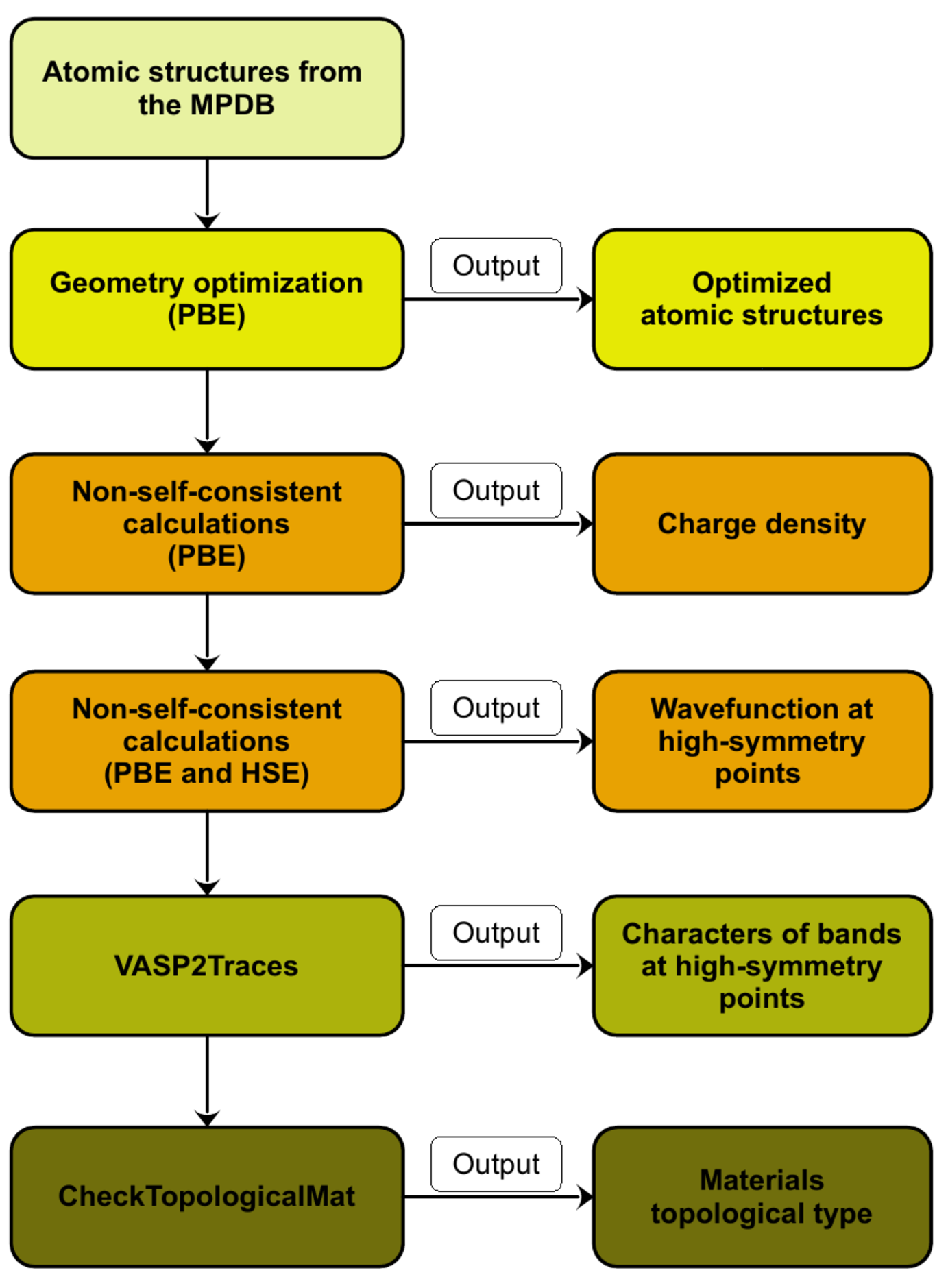}
\caption{The workflow for predicting the topological nature of materials.}
\label{fig:wf}
\end{figure}

Following the optimization step, the atomic structures are then used for non-self-consistent calculations with PBE to compute the charge density. In the subsequent step, the wavefunctions at specific high-symmetry points in the Brillouin zone are calculated with both the PBE and HSE functionals. The wavefunctions are processed with the VASP2Trace code~\cite{Vergniory2019}, which computes the symmetry operators and the coefficients of the plane waves at the high-symmetry points. The trace files generated by VASP2Trace contain the eigenvalues at each maximal k-vector (high-symmetry points) of a space group. Finally, the CheckTopologicalMat tool analyzes these data to classify materials into their respective topological types, such as trivial insulator, topological insulator, or semimetal, completing the systematic classification process.

\begin{figure}[ht]
\centering
\includegraphics[width=0.49\linewidth]{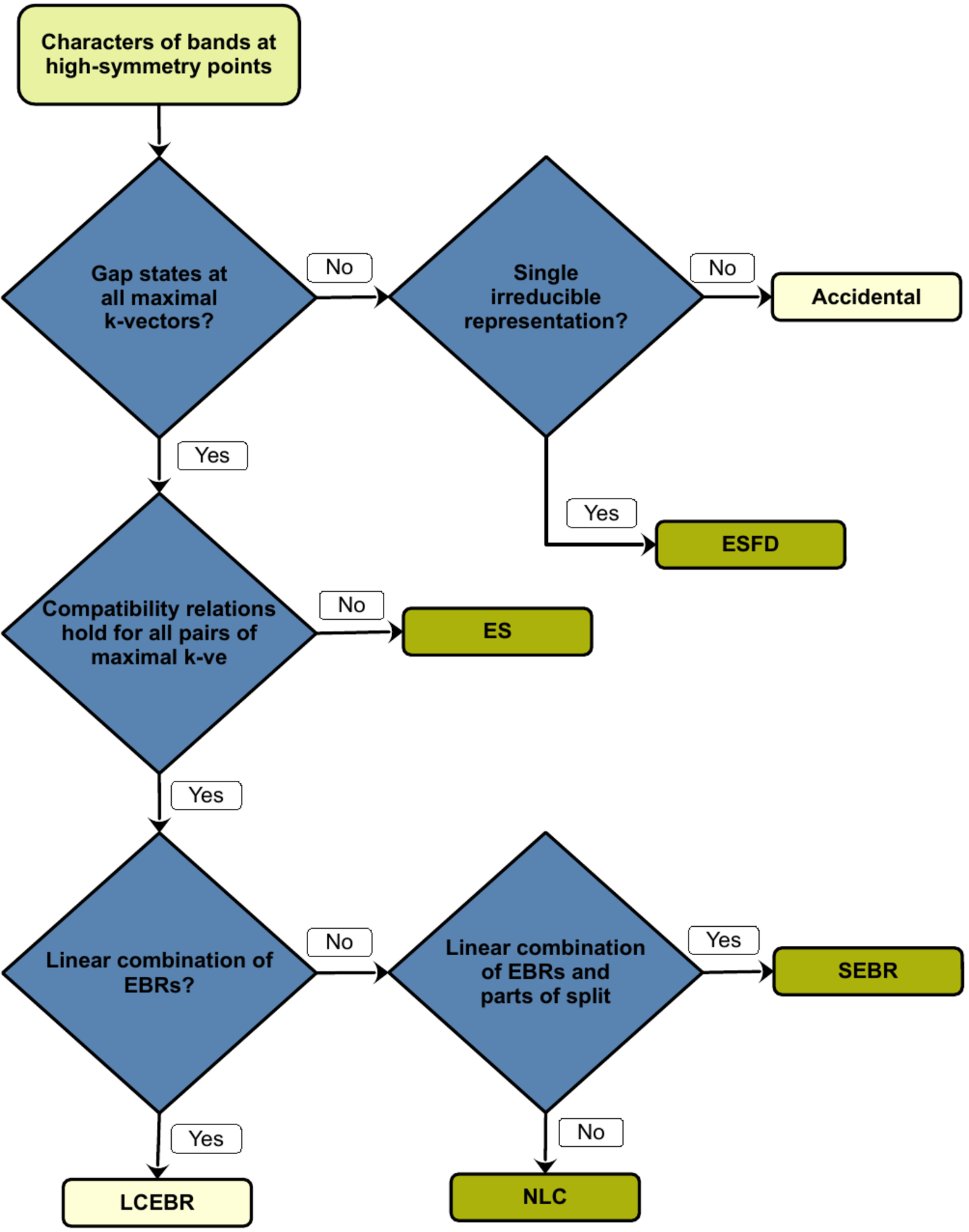}
\caption{The CheckTopologicalMat flowchart.}
\label{fig:ctm}
\end{figure}

The CheckTopologicalMat tool, available at the Bilbao Crystallographic Server (https://www.cryst.ehu.es) uses the compatibility relations and the set of elementary band representations (EBRs) to classify the materials. The process of the material classification by CheckTopologicalMat is summarized in the flowchart shown in Figure~\ref{fig:ctm}. The initial step is to determine whether gapped states exist at all maximal k-vectors (the special high-symmetry points in the Brillouin zone). In the absence of gapped states, the subsequent step involves verifying whether the bands form a single irreducible representation. Materials are then classified into two distinct categories: ``Enforced Semimetals with Fermi Degeneracy'' (ESFD), which are semimetals with degeneracies enforced by symmetry at the Fermi level, and ``Accidental'', where band crossings occur due to accidental degeneracy rather than being protected by symmetry.

In the event of gapped states at all maximal k-vectors, the tool evaluates the compatibility relations between pairs of maximal k-vectors in the Brillouin zone. If the compatibility relations fail for any pair, then the material is classified as an ``Enforced Semimetal'' (ES), where band crossings are symmetry protected but may not necessarily occur at the Fermi level. Conversely, if the compatibility relations are satisfied, the band structure is analyzed to determine whether it can be expressed as a linear combination of EBRs. A band structure that matches EBRs exactly corresponds to linear combination of EBRs (LCEBR), indicating a trivial material with no topological characteristics. If the band structure partially matches EBRs (combination of full and split EBRs), then the material is classified as ``Split Elementary Band Representations'' (SEBR). If no combination of EBRs or split EBRs fits, the material is classified as ``No Linear Combination'' (NLC).

Our findings are detailed in Tables~\ref{tab:nmbrs_1} and~\ref{tab:nmbrs_2}. Analyzing 12,035 electronic structures calculated by PBE, we identified 554 materials classified as NLC and 796 as SEBR. Additionally, 1,158 electronic structures were categorized as ESFD, and 912 as ES. These results reveal that more than 28\% of the PBE electronic structures exhibit topologically nontrivial properties. This is in agreement with previous studies reporting that 27-30\% of materials exhibit topological properties~\cite{Vergniory2019, Zhang2019}. It should be noted that this consistency is achieved despite the fact that our study has a smaller sample size compared to previous studies. 

% Andreas: we should have a new paragraph here

Notably, the results differ when the band structures are calculated using the HSE hybrid functional, underscoring the sensitivity of topological classification to the choice of the XC functional. Specifically, of the 9,757 electronic structures calculated with HSE, 298 are categorized as NLC, 407 as SEBR, 371 as ESFD, and 378 as ES. That is, approximately 15\% of the materials are identified as TMs (see Table~\ref{tab:nmbrs_1}). 
\begin{table}[ht]
\centering
\begin{tabular}{|l|l|l|l|l|l|}
\hline
 & Total & NLC  & SEBR & ESFD & ES \\
\hline
PBE & 12035 & 554 (4.6\%) & 796 (6.6\%) & 1158 (9.6\%) & 912 (7.6\%) \\
\hline
HSE & 9757  & 298 (3.0\%) & 407 (4.2\%) & 371 (3.8\%)  & 378 (3.9\%) \\
\hline
\end{tabular}
\caption{\label{tab:nmbrs_1} Number of TMs for the band structures calculated with PBE or HSE.}
\end{table}

A total of 9,571 electronic structures are classified by both the PBE and HSE functionals. The results are detailed in Table~\ref{tab:nmbrs_2}. Among the PBE band structures, we identified 866 topological insulators (NLC and SEBR) and 787 topological semimetals (ES and ESFD). In comparison, 671 of the HSE band structures are topological insulators (NLC and SEBR) and 703 are topological semimetals (ES and ESFD). Notably, the majority of materials identified as ESFD by HSE (334 out of 346, or 96\%), retain the same classification with PBE. The exceptions are 12 materials, which are categorized as ES (6), SEBR (5), and trivial insulator or metal (1). 
%\textcolor{red}{all 5 SEBR compounds have the following in common: possible transformations ESFD -> TI lowering the symmetry.}
ES has a similar trend, where of the 357 materials classified as ES by HSE, 314 (88\%) maintain their classification with PBE, with 2 SEFD, 4 NLC, and 30 SEBR. 7 materials are classified as trivial insulator or metal. 

\begin{table}[t]
    \centering
    \begin{tabular}{|l|l|l|l|l|l|l|}
    \hline
     & NLC  & SEBR & ESFD & ES & Trivial & Total\\
    \hline
     NLC  & 255 & 11 & 0 & 7 & 8 & 281 \\
    \hline
    SEBR & 11 & 345 & 0 & 23 & 11 & 390 \\
    \hline
    ESFD & 0 & 5 & 334 & 6 & 1 & 346 \\
    \hline
    ES & 4 & 30 & 2 & 314 & 7 & 357 \\
    \hline
    Trivial & 100 & 105 & 32 & 69 & 7892 & 8197 \\
    \hline
    Total & 370 & 496 & 368 & 419 & 7918 & 9571 \\
    \hline
    \end{tabular}
    
\centering
\caption{\label{tab:nmbrs_2} Number of TMs for the band structures calculated with PBE and HSE. The rows show the corresponding numbers for HSE, while the columns show the corresponding numbers for PBE.}
\end{table}

The consistency between HSE and PBE in the classification of topological semimetals can be understood in the context of how the topological nature of semimetals is determined. In crystalline solids, electronic bands must satisfy specific symmetry constraints dictated by the compatibility relations. When bands fail to satisfy these compatibility relations at certain high-symmetry points or along high-symmetry lines, the resulting band structure may exhibit features such as band crossings that are not coincidental and protected by the crystal's symmetry. These crossings are typically associated with Dirac or Weyl points. ES materials exhibit a band structure that features a protected crossing of conduction and valence bands at certain points in the Brillouin zone. ESFD materials are a special subclass of ESs where the Fermi level lies exactly at the energy of the band-crossing points. The discrepancy between HSE and PBE results arises from differences in how each method calculates band structures/band filling.

Regarding topological insulators, 26 out of 281 materials (about 9\%) classified as NLC with HSE are classified differently with PBE, specifically 7 ES, 11 SEBR, and 8 trivial insulator or metal. For the SEBR class, out of 390 materials classified as SEBR by HSE, 345 (88\%) retain the same topological classification with PBE. The exceptions are 45 materials, which are categorized as ES (23 materials), NLC (11 materials), and trivial insulator or metal (11 materials). Topological insulators are characterized by electronic states that satisfy the compatibility relations across the Brillouin zone. If the valence bands of an insulator cannot be expressed as a sum of EBRs, the material is considered topological. The classification of these materials is based on how band structure can be decomposed into EBRs~\cite{Bradlyn2017}. 
 The band structure of SEBR materials shares similarities with that of graphene, where an EBR splits into disconnected valence and conduction bands. In these materials, the electronic bands just below and just above the Fermi level together form an EBR. In NLC electronic structures the bands below and above the Fermi level collectively form two EBRs. In these materials, neither the valence nor the conduction bands belong to a split EBR. The discrepancy between the PBE and HSE classifications stems from  differences in the PBE and HSE band structures. HSE results are considered superior due to the ability of the HSE functional to predict the bandgap of materials more accurately than the PBE functional.

\begin{figure}[p]
\centering
\includegraphics[width=0.45\linewidth]{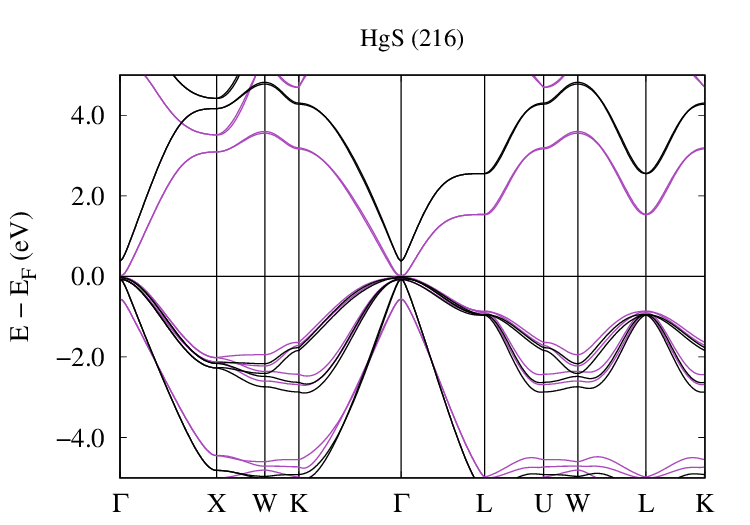}
\includegraphics[width=0.45\linewidth]{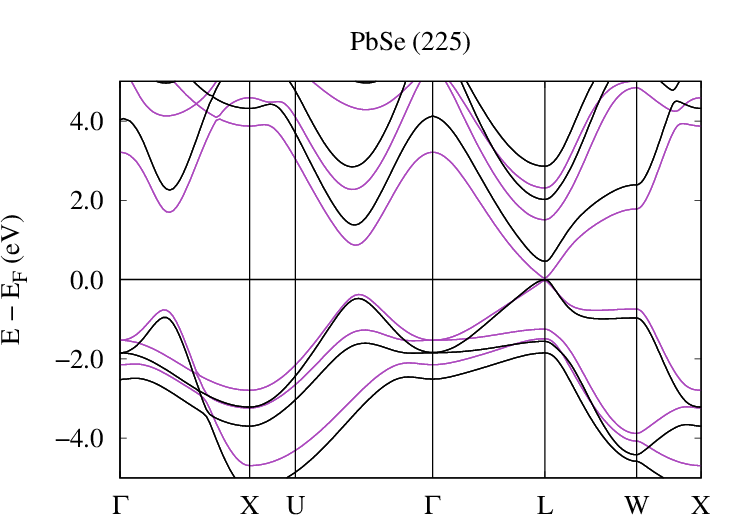}
\includegraphics[width=0.45\linewidth]{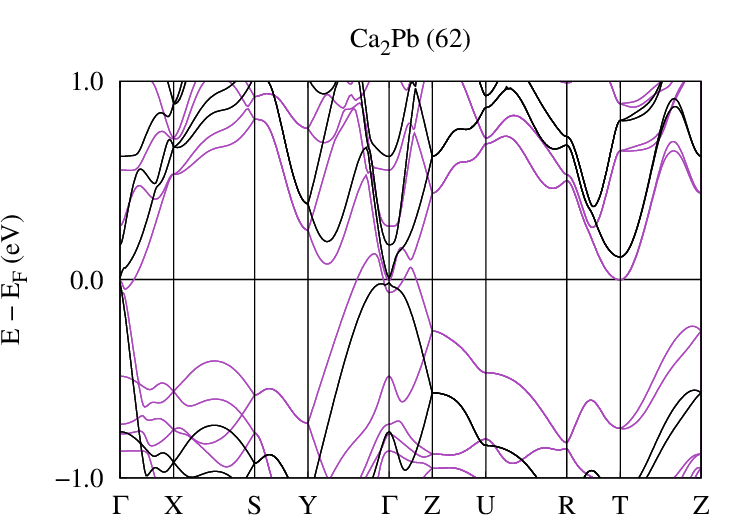}
\includegraphics[width=0.45\linewidth]{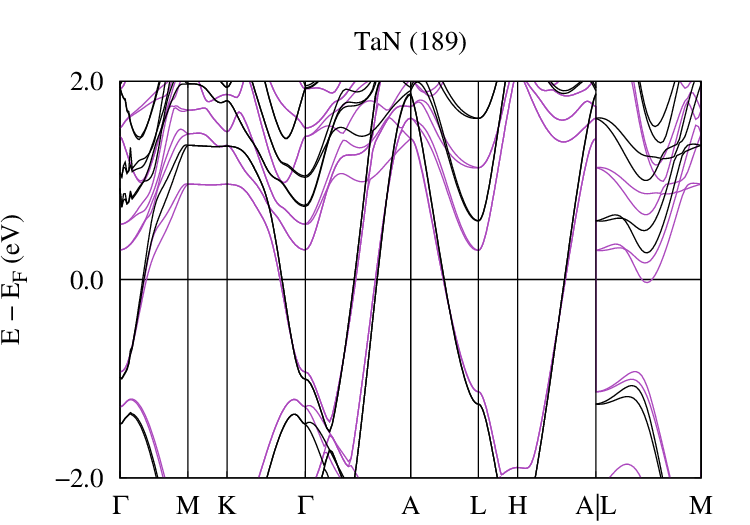}
\includegraphics[width=0.45\linewidth]{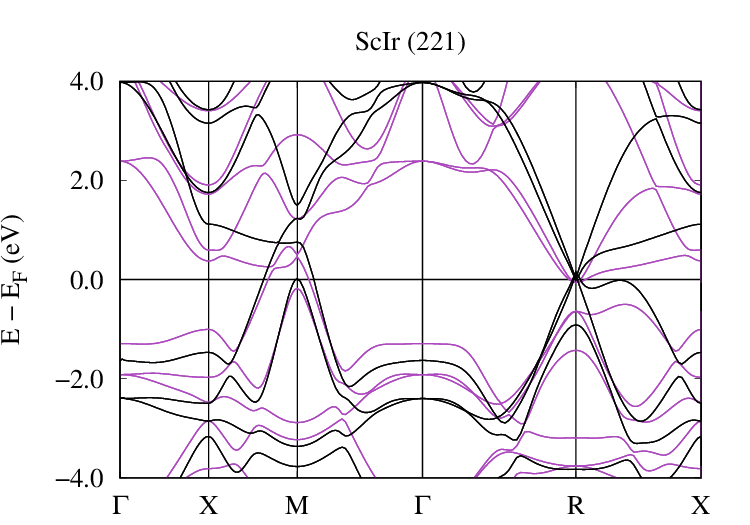}
\includegraphics[width=0.45\linewidth]{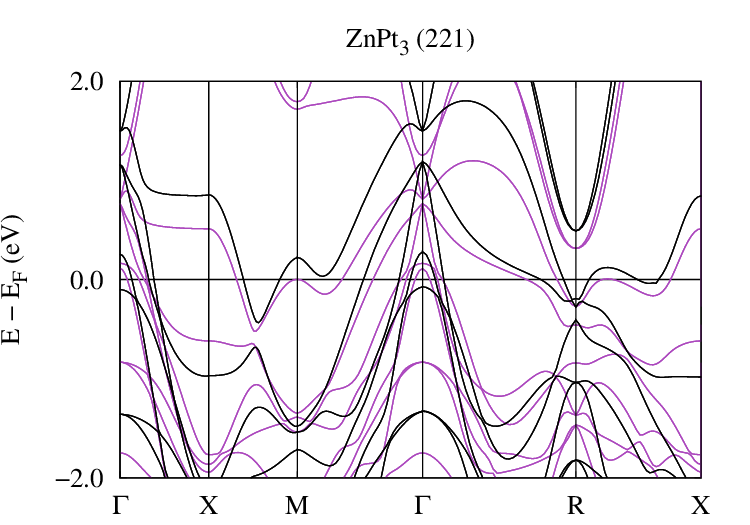}
\includegraphics[width=0.45\linewidth]{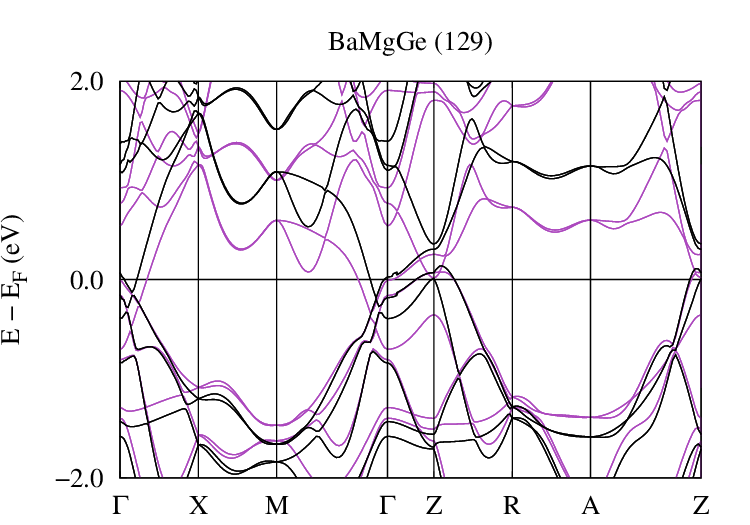}
\caption{Band structures of selected materials calculated with PBE (purple) and HSE (black). }
\label{fig:bs}
\end{figure}

The band structures of selected materials, calculated by PBE and HSE, are presented in Figure~\ref{fig:bs}. The HSE band structures of HgS and PbSe exhibit a bandgap that is absent in the PBE band structures. HgS and PbSe crystallize in the cubic system with space groups F\={4}3m (216) and Fm\={3}m, respectively. The changes in the band structures of these compounds are small, except for the opening of the gap in their HSE band structures. However, the topological classification has changed. Both materials are classified as trivial insulators when evaluated with HSE. Consistent with previous reports~\cite{Vergniory2017}, the PBE band structure of HgS is classified as NLC, whereas PbSe is categorized as a trivial insulator. $\text{Ca}_\text{2}\text{Pb}$, which crystallizes in the orthorhombic (Pnma), shows a small bandgap in its HSE band structure and is classified NLC. The PBE band structure of $\text{Ca}_\text{2}\text{Pb}$ shows band crossing and is categorized as trivial. The differences between the HSE and PBE electronic structures of TaN, which crystallizes in the hexagonal (P\={6}2m), is negligible. However, a modification in the nature of bands at specific k points results in a change in the topological classification. TaN is identified as ES with PBE, in agreement with previous studies, and as SEBR when using HSE. ScIr and $\text{ZnPt}_\text{3}$ crystallize in the cubic system with space groups Pm\={3}m. Several bands cross the Fermi level indicating both compounds are metal. While HSE calculations identified ScIr as SEBR, its PBE band structure is categorized as NLC. The HSE band structure of $\text{ZnPt}_\text{3}$ is identified as ES, whereas the band structures calculated with PBE is classified as SEBR. BaMgGe (tetragonal, P4/nmm1) is classified as NLC with HSE and ES with the PBE functional.

%\textcolor{red}{
%mp-1123 HgS  NLC (NLC) --> LCEBR \newline
%mp-2201 PbSe LCEBR (LCEBR/SEBR) --> LCEBR \newline
%mp-30478 Ca2Pb LCEBR (LCEBR) --> NLC \newline
%mp-1129 ScIr NLC (NLC) --> SEBR \newline
%mp-1279 TaN  ES (ES) --> SEBR \newline
%mp-754515 BaMgGe ES (NLC) --> NLC \newline
%mp-30856-ZnPt3 SEBR (SEBR) --> ES \newline
%}
%

Despite the exciting activity in identifying TMs, the search for TMs suitable for real-world applications remains a key bottleneck. The identification of TMs through first-principles calculations represents a crucial initial step in the design of materials for practical use. Predicting the topological behavior of materials using hybrid functionals, renowned for their accuracy in describing electronic structures, serves as an essential step for high-throughput materials screening.

In summary, we investigated the classification of TMs by employing enhanced computational workflows. Our study refines conventional approaches by optimizing atomic structures from the MPDB and utilizing the HSE hybrid functional for electronic structure calculations. A dataset of 12,035 non-magnetic materials was analyzed using PBE and HSE functionals to determine topological properties. Results reveal that 28\% of PBE-based calculations classify materials as topologically nontrivial. HSE calculations indicate only 15\% of materials exhibit topological properties, demonstrating the critical influence of band structure calculations on topological classification.

\section*{Methods}
Our workflows are developed using atomate\cite{atomate}, which is built on top of open-source Python libraries such as pymatgen\cite{pymatgen} and FireWorks\cite{fireworks}. These tools provide workflow templates for the automated computation of various materials properties, including electronic structure. Electronic structure calculations were carried out within the framework of DFT, employing the projector augmented wave (PAW) method \cite{PhysRevB.50.17953}, as implemented in the Vienna Ab initio Simulation Package (VASP)\cite{Kresse1996_1}.  

\begin{figure}[!h]
\centering
\setlength{\fboxsep}{0pt}%
\setlength{\fboxrule}{0.1pt}%
\fbox{\includegraphics[width=0.7\linewidth]{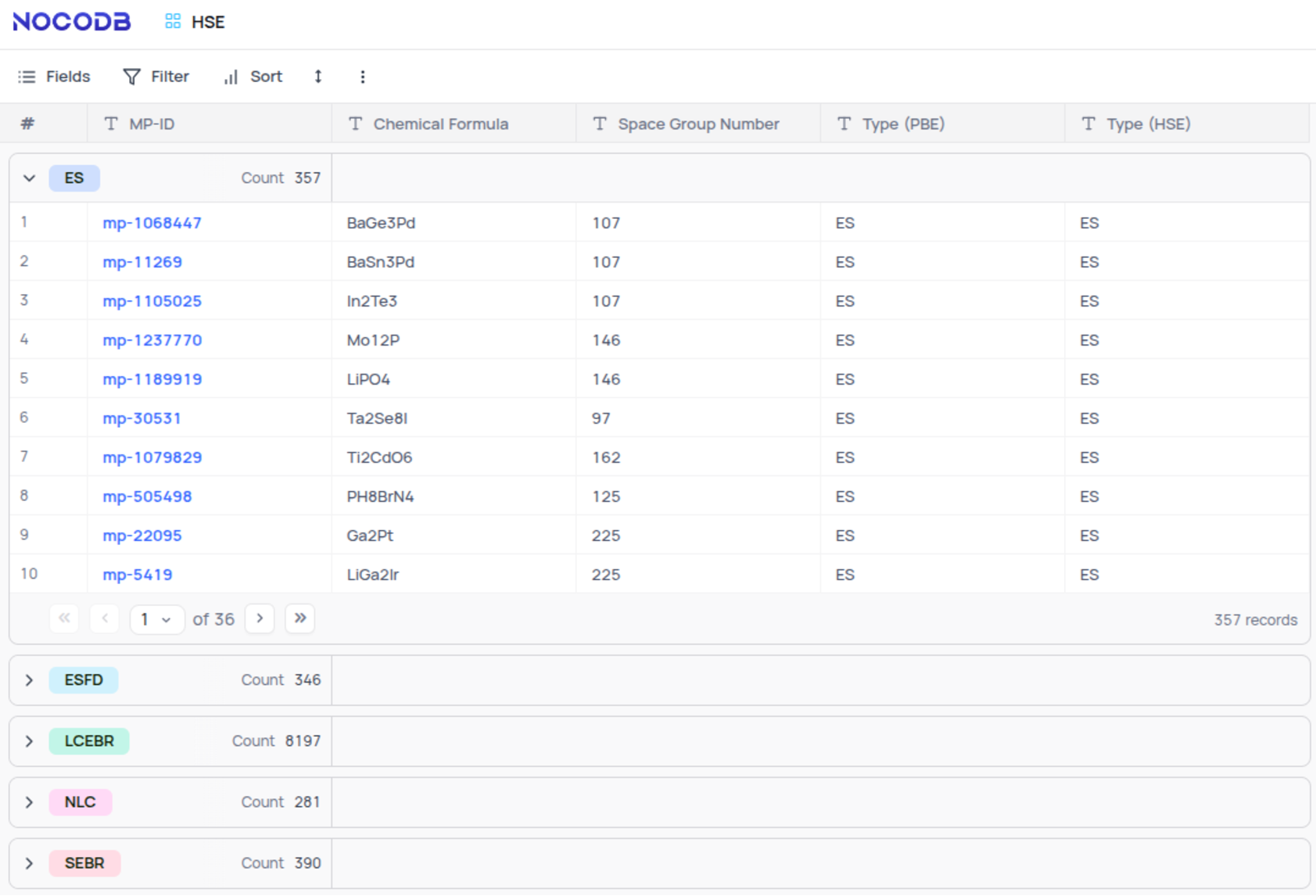}}
\caption{A screenshot of our data inventory showing the entries grouped by their HSE topological type.}
\label{fig:data}
\end{figure}

Our workflow begins by retrieving atomic configurations which contain fewer than 31 atoms per unit cell from the MPDB. As of July 2022, the MPDB comprises 49,794 experimentally observed materials, 30,965 of which contain fewer than 31 atoms per unit cell. Materials containing V, Cr, Mn, Fe, Co, Ni, lanthanide, and actinide series were excluded from our list. The remaining 14,288 compounds were subsequently subjected to geometry optimization.

Geometry optimization was performed using the Perdew-Burke-Ernzerhof (PBE) functional~\cite{PhysRevLett.77.3865}, with a plane-wave energy cutoff set to 520 eV. The smallest allowed spacing between k-points (KSPACING) was set to 0.30~\AA$^{-1}$. Atomic structures were considered relaxed when the norm of all forces fell below 10~meV/\AA. In line with previous studies, the $paw\_PBE.52$ pseudopotentials were employed. Compounds with an odd number of electrons per unit cell were excluded from the workflow. In this step, 12,607 atomic configurations were successfully optimized.

For the electronic structure calculations, we first performed non-self-consistent calculations using PBE. Subsequently, for each compound, the wavefunctions at selected high-symmetry points in the Brillouin zone were computed using both the PBE and HSE functionals. We have used HSE06~\cite{Krukau2026} which uses a screening parameter of 0.11 $\text{\AA}^{-1}$ and 25\% exact Hartree-Fock exchange for short-range interactions.
Spin–orbit coupling (SOC) was incorporated self-consistently via the second variation method~\cite{Hobbs2000}. During this step, compounds with magnetic moments exceeding 0.1 $\mu_{B}$ per atom were identified as magnetic and excluded from further analysis. For the remaining non-magnetic compounds, symmetry operators and plane-wave coefficients at high-symmetry points were computed and stored in ``trace'' files. These files were then processed to extract information regarding the topological properties of the materials.

Our dataset inventory uses the \href{https://nocodb.com/}{NOCODB} project, which offers enhanced, spreadsheet-like functionality within a web browser including convenient searching, filtering, sorting, and grouping of data. It also supports the creation and management of multiple predefined views, which can be shared publicly or confidentially. The data can be downloaded in CSV and Excel formats. A screenshot of our dataset inventory is provided in Figure~\ref{fig:data}. Each entry includes the chemical formula, space group number, the topological classifications according to PBE and HSE, and is hyperlinked to its corresponding page in the MPDB~\cite{Anubhav2013}.  All metadata can be filtered and sorted in CSV and Excel formats for convenient browsing. Our publication of the data and the metedata addresses the F.A.I.R. data principles~\cite{Wilkinson2016}, in particular the F3, F4, A1, I1, and I3 criteria.

One of the key aspects of NOCODB is its emphasis on a "no-code" approach, making it accessible to users without programming expertise. Despite this, it offers powerful and robust REST APIs and integrations with a wide range of established web services. This enables advanced automation and customization for users with coding experience. Here, it was used for automated data ingest from the computational results.
NOCODB operates on top of relational databases such as MySQL, PostgreSQL, or SQLite, ensuring scalability to handle extensive metadata catalogs with thousands of columns and millions of rows. For this project, we use the self-hosted, open-source version of NOCODB.
 
\section*{Data availability}
All raw data including trace files and outputs of CheckTopologicalMat are available at the Rossendorf Data Repository (RODARE) https://rodare.hzdr.de/. In addition, material classifications based on PBE and HSE electronic structure calculations are publicly available at https://data.casus.science/dashboard/\#/nc/view/97e10cab-d273-4cf8-985b-3bc2340b4ae3.

\bibliography{tms}

\section*{Acknowledgements}
The authors gratefully acknowledge the computing time provided to them on the high-performance computer Noctua2 at the NHR Center PC2. This is funded by the Federal Ministry of Education and Research and the state governments participating on the basis of the resolutions of the GWK for the national high-performance computing at universities (www.nhr-verein.de/unsere-partner). The computations for this research were performed using computing resources under project hpc-prf-abtop.
The authors gratefully acknowledge Maia Garcia Vergnior and Nicolas Regnault for the fruitful discussions that contributed to the development of this work. 

\section*{Author contributions statement}

H.M. conceived the idea of the paper and performed all DFT calculations.
\\
L.E. analyzed the band structures and contributed to the writing of the manuscript.  
\\
A.K. created the metadata inventory and contributed to the writing of the manuscript.
\\
T.D.K. contributed to the writing of the manuscript.
\\
All authors reviewed the manuscript. 

\section*{Additional information}
The author(s) declare no competing interests.

\end{document}